\documentclass{article}
\usepackage{spconf,amsmath,epsfig}
\usepackage{tabularx}
\usepackage{booktabs}
\usepackage{enumitem}
\usepackage{float}
\setlist[itemize]{noitemsep, topsep=0.5em, leftmargin=*, labelsep=0.7em}
\usepackage{xcolor}
\usepackage{subcaption}
\usepackage{listings}
\let\OLDthebibliography\thebibliography
\renewcommand\thebibliography[1]{
  \OLDthebibliography{#1}
  \setlength{\parskip}{0pt}
  \setlength{\itemsep}{0pt plus 0.3ex}
}

\usepackage{fancyhdr}
\fancypagestyle{plain}{
  \fancyhf{}
  
  \fancyhead[C]{IEEE ICME Workshop on Coding for Machines, Brisbane, Australia, 2023.}
}
\usepackage[margin = 1.9cm, headsep=0.3cm,headheight=1cm,]{geometry}
\usepackage{caption}

\expandafter\def\expandafter\normalsize\expandafter{%
    \normalsize
    \setlength\abovedisplayskip{5pt}
    \setlength\belowdisplayskip{7pt}
    \setlength\abovedisplayshortskip{5pt}
    \setlength\belowdisplayshortskip{7pt}
}
\usepackage{xpatch}

\begin{document}\sloppy

\def\x{{\mathbf x}}
\def\L{{\cal L}}

\title{VVC+M: Plug and Play Scalable Image Coding for Humans and Machines}
%
\name{Alon Harell, Yalda Foroutan, and Ivan V. Baji\'{c} 
\thanks{\copyright~2023 IEEE. Personal use of this material is permitted. Permission from IEEE must be obtained for all other uses, in any current or future media, including reprinting/republishing this material for advertising or promotional purposes, creating new collective works, for resale or redistribution to servers or lists, or reuse of any copyrighted component of this work in other works.}}

\address{School of Engineering Science, Simon Fraser University, Burnaby, BC, Canada}

\maketitle

\begin{abstract}
Compression for machines is an emerging field, where inputs are encoded while optimizing the performance of downstream automated analysis. In scalable coding for humans and machines, the compressed representation used for machines is further utilized to enable input reconstruction. Often performed by jointly optimizing the compression scheme for both machine task and human perception, this results in sub-optimal rate-distortion (RD) performance for the machine side. We focus on the case of images, proposing to utilize the pre-existing residual coding capabilities of video codecs such as VVC to create a scalable codec from any image compression for machines (ICM) scheme. 
Using our approach we improve an existing scalable codec to achieve superior RD performance on the machine task, while remaining competitive for human perception. Moreover, our approach can be trained post-hoc for any given ICM scheme, and without creating a coupling between the quality of the machine analysis and human vision. 
\end{abstract}
%
\section{Introduction}

\label{sec:intro}

Recent trends in computer vision (CV) have seen an increase in the bandwidth used for communication of images and video for processing by automated task models. As a result, the emerging field of compression for machines (sometimes referred to as coding for machines, CM) has garnered growing attention from both the computer vision and compression communities, including standardization efforts by leading bodies~\cite{mpeg_vcm}. In lossy coding, an input is encoded and later reconstructed imperfectly, leading to some degradation, known as distortion. The goal of traditional lossy codecs is to minimize the size of the encoding, known as rate, while incurring as little distortion possible in terms of the human perception of the reconstructed 
image. In compression for machines, however, human perception is no longer a priority, and the distortion is measured in terms of the performance of the automated CV models, performed on the decoded representation. Most CV tasks require less information to be completed successfully than what is present in the full image or video, meaning that CM codecs can often achieve better rate-distortion (RD) performance than traditional compression.

Although CM offers significant rate savings, in some applications it may be necessary to reconstruct the input fully on occasion. For example, consider an automated traffic camera, which uses a server-side task model to detect cars, identify license plates and monitor red-light violations. Whenever a violation is committed, the raw footage may be required as evidence in any legal proceedings. Recent works~\cite{hyomin, pcs, ozyilkan2023learned, semantics_scalable, face_scalable}, offer a solution to such scenarios that balances the rate-distortion performance of both the CV task and full input reconstruction. In the scalable coding setting, the encoding is performed in layers - in the first, ``base-layer," information needed for automated analysis is encoded and subsequently decoded and utilized to perform the CV task; in the next, ``enhancement-layer"\footnote{In some cases, such as~\cite{hyomin, semantics_scalable}, there may be multiple enhancement-layers, corresponding to growing CV task complexity, or growing task accuracy} additional information is used alongside the base-layer encoding to fully reconstruct the image. 

\begin{figure*}[t!]
  \centering
  \includegraphics[width=0.98\linewidth]{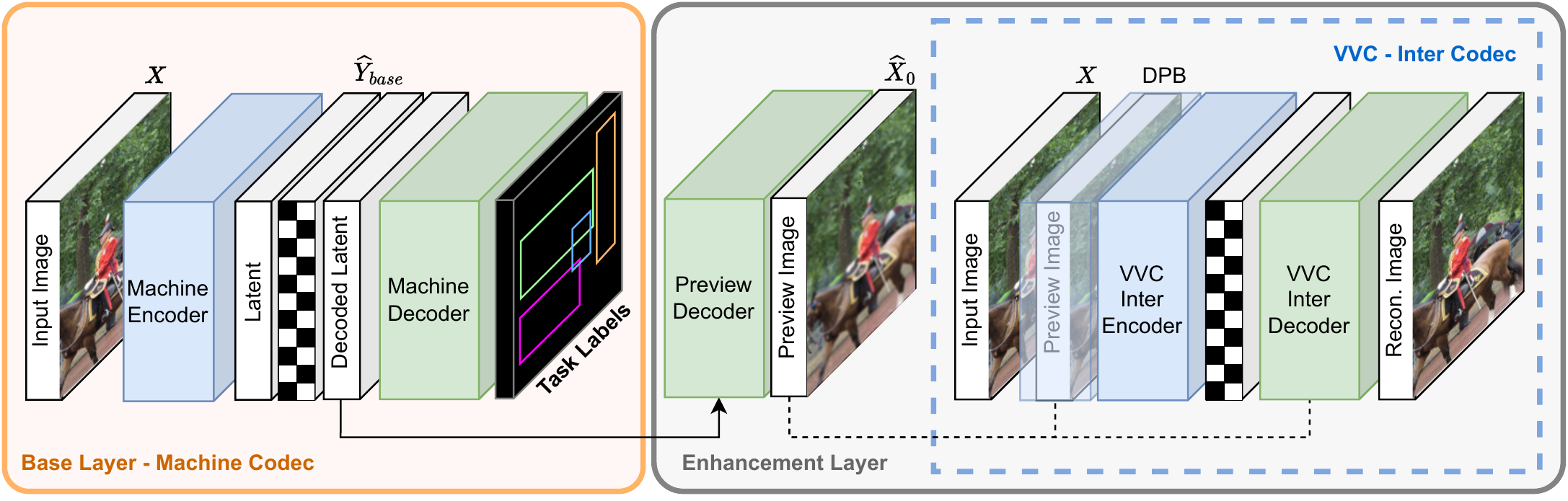}
  \caption{Proposed scalable coding approach - Checkerboard blocks represent the arithmetic encoding and decoding, and correspond to the base and enhancement bitstreams. The preview codec is trained to recreate a preview image $\widehat{X}_0$ from the latent representation of the machine codec (which can be trained completely separately). The preview image is then inserted into the decoded picture buffer (DPB) of a VVC codec to allow for efficient inter-prediction of the original image. 
  }
  \label{fig:prop}
\end{figure*}

In learned compression, a traditional, hand-crafted, compression scheme is replaced with a machine-learning model, often a deep neural network, trained end-to-end to optimize a rate-distortion objective of the form:
\begin{equation}\label{eq:learned}
    L = R + \lambda\cdot D.
\end{equation} 
Here, $R$ is 
a loss term corresponding to the rate and $D$ is a distortion loss, while $\lambda$ is 
a Lagrange multiplier balancing the two. Recently, learned compression methods have greatly improved in performance and popularity, achieving comparable and sometimes even superior performance to traditional codecs both for images~\cite{balle2018variational,akbari2019dsslic, minnen2018joint, cheng2020image, he2022elic}, as well as videos~\cite{DVC2019_CVPR, ho2022canf}. 
The learnable compression approach lends itself naturally to both CM as well as scalable coding for humans and machines, since we simply have to adjust the terms $R$ and $D$ for the desired setting. Notably, learned scalable codecs are often trained jointly, meaning a single model (comprised of the aforementioned two layers) is trained with a single objective, balancing the optimization of both layers. 

Although the joint training has some advantages, such as creating a base representation which is easy to use for the enhancement task, it is not without downsides. Most importantly, it may lead to an overly inflated base representation, which contains some information not directly utilized by the task model. In scenarios where the base-layer is needed significantly more often than enhancement (such as our traffic camera), this leads to an overall degradation in rate-distortion performance. An important downside of many of the current methods for learned compression is that they require re-training the model for each desired quality (and thus bitrate) level. This means that the joint training of scalable models also introduces a potentially unwanted coupling between the qualities of the machine analysis task and human perception.

In~\cite{akbari2019dsslic}, the authors create a scalable coding for varying quality of human vision by utilizing a segmentation model on the encoder side as side information. While this leads to some benefits in input reconstruction, it cannot be considered a scalable model for humans and machines, because the entire machine task is performed on the encoder side, and only it's corresponding output is encoded. In~\cite{semantics_scalable, face_scalable}, the base representation is used to obtain a low-quality reconstruction of the input, sometimes known as a \emph{preview}. The preview is then subtracted from original input to create a residual which is encoded using BPG, the intra-coding method used in in HEVC~\cite{hevc_std_2019}. An important downside of this approach is that BPG is optimized for regular images, not residual images, leading to potential sub-optimality in this setting. 

In this paper we propose an improved method for residual coding 
in scalable human-machine systems by utilizing \textbf{inter-prediction coding} 
from a video codec. Inter-prediction is used in many video codecs to encode an image (frame) based on one or more previously decoded frames and thus is far better suited for residual coding. Our main contributions are:
\begin{itemize}
    \item Improve the machine-task RD performance of scalable human-machine image codecs by separately optimizing base-layer for task performance only, which allows for close or superior task RD performance compared with current state-of-the-art (SOTA)  for object detection.  
    \item Propose a method for transforming any 
    ICM codec to a scalable human-machine codec using residual coding modes of a video codec.
    \item Utilize the existing powerful VVC codec to design our enhancement-layer, achieving competitive RD performance for human perception with minimal training.
    \item Derive a methodology for comparison between two scalable human-machine codecs 
    based on Bj\o{}ntegaard metrics~\cite{Bjontegaard}, which takes both base and enhancement layers into account. Using this metric we demonstrate that our approach outperforms current 
    SOTA scalable human-machine codecs in many relevant scenarios. 
\end{itemize}
The rest of the paper is presented as follows: Section~\ref{sec:proposed} presents an overview of our proposed method, including detail about our base and enhancement codecs; Section~\ref{sec:experimental} details our experimental setup, presents results of both layers of our scalable compression approach, including our proposed comparison metric; finally, Section~\ref{sec:summary} provides discussion and summary of our work.

\section{Proposed Method}\label{sec:proposed}

Our proposed scalable human-machine image 
codec, which is visualised in Fig.~\ref{fig:prop}, 
is comprised of 3 parts: a base-layer performing object detection, a preview decoder, and an enhancement-layer. For the base-layer, we can use \textbf{any ICM codec}, regardless of the task it performs or the method it employs to do so, as long as we have access to its decoded representation, 
denoted $\widehat{Y}_{base}$. 
We train a preview synthesis model $g_s^{prev}$ to create a preview image, $\widehat{X}_0 = g_s^{prev}(\widehat{Y}_{base})$, from the latent base representation. Finally, we utilize an existing video codec to efficiently encode the input $X$ using our preview image $\widehat{X}_0$ as a reference, in a process commonly known as inter-coding. We use the VVC~\cite{vvc_std} codec, as implemented in VTM version 12.3~\cite{VTM12.3} to perform the inter-coding, although generally any video codec can be used. 

\subsection{Base Layer}\label{subsec:proposed_base}

An important advantage to our approach is that it allows us to separately optimize the base-layer, avoiding the disadvantages of joint training, as explained in the introduction. Even though our scalable approach can utilize any ICM codec as a base-layer, we implement our base-layer to resemble~\cite{hyomin}, which until recently achieved state-of-the-art RD performance for object detection. We do this by using 
the same encoder structure, and 
discarding the enhancement representation (which is done by reducing the dimensions of the final layer of the encoder), while 
keeping the base-layer decoder the same as in
~\cite{hyomin}. For more details, see~\cite{hyomin}, or 
Appendix~\ref{app:base_layer} of the supplemental material. Of course, unlike~\cite{hyomin}, we train the base-layer codec for the base task only, instead of joint training with the enhancement-layer. Using a nearly identical base-layer implementation allows to directly showcase the advantages of training the base-layer to solely optimize the RD performance of the base task, and avoids confounding these effects with the differences between base-layer models. 

\subsection{Enhancement Layer}\label{subsec:proposed_enh}

In order to allow for maximum flexibility, we construct our enhancement-layer so that is agnostic to the base-layer structure. In fact, the only input needed for our enhancement-layer is the decoded base-bitstream after arithmetic decoding, which in our case is denoted $\widehat{Y}_{base}$. First, we use a synthesis transform $g_s^{prev}$ to obtain a 
preview of the input. We structure this preview synthesis model similarly to the decoder of~\cite{cheng2020image}, which is comprised of residual convolutional blocks, upsampling blocks and sub-pixel convolutions, using a combination of leaky ReLU and inverse GDN activations.  

In the next step, the preview image $\widehat{X}_0$ is inserted into the \emph{decoded picture buffer} (DPB) of a VVC codec, where it is used to encode the original image $X$ using inter-coding. 
Using currently available implementations of VVC, inserting an image directly into the picture buffer is not possible, leading us to take the following approach insted. At first, the preview and input image are converted into YUV444 format using FFMPEG~\cite{ffmpeg}, and joined to create an uncompressed two-frame video sequence. This video is then encoded and decoded using VVC reference software VTM (version 12.3)~\cite{VTM12.3} set to 
the low-delay-P coding setting. Because the preview image is based on the base-layer representation, which is already available on the decoder side, it would not need to be encoded (or decoded) in a practical setting. To account for this, we must ensure that the preview is encoded with minimal or no loss, without regard to its bitrate. In order to ensure the preview image does not suffer further degradation, we set it to be an intra-frame\footnote{Intra-frames are encoded and decoded without the help of any other frames in the video, generally requiring the most bits. Additionally, intra-frames are also commonly encoded at a higher quality then other frames to ensure they make for a reliable reference.} and use the \texttt{IntraQPOffset} setting of VTM to encode it without loss (equivalent to \texttt{QP = 0}), regardless of the desired quality of the enhancement image. 
Of course, since the process of encoding the preview frame is only used as a workaround for implementation reasons, we only count the bit-rate used for the coding of the second frame (which is reported separately by VTM), when evaluating our enhancement-layer. An example of a full configuration file for VTM for our setup is provided in Appendix~\ref{app:cfg} of the supplemental material. 

\section{Experiments}\label{sec:experimental}
As explained in our proposed method, the training of our base and enhancement layers can be done separately. Furthermore, the only trainable portion of our enhancement-layer is the preview synthesis transform, which is quick and relatively simple to train. We begin by training two versions of our base-layer corresponding to two popular DNN models for object detection: YOLOv3~\cite{Redmon2018_yolov3} and Faster R-CNN~\cite{ren2016faster}. We pick these models because they are well-established, well-understood models and because they were used in previously set benchmarks for scalable coding for humans and machines~\cite{hyomin,pcs, ozyilkan2023learned}. Training is performed using the Lagrangian rate distortion loss shown in Eq.~\ref{eq:learned}, where the distortion is simply the mean squared error (MSE) with respect to the same feature layer $F$. We choose the same features as~\cite{hyomin}, and vary the value of $\lambda$ to get different points on the RD curve. In the case of Faster R-CNN, as was the case in ~\cite{hyomin}, the feature layer $F$ is comprised of a concatenation of several tensors, and an average of their MSE was used as an objective.  

All models were trained in a similar approach to~\cite{hyomin,pcs}, using two-stage training. At the first stage (300 epochs for YOLOv3, 500 epochs for Faster R-CNN), randomly cropped patches of size $256\times256$ are taken from a combination of the JPEG-AI~\cite{jpeg_ai_dataset} and CLIC~\cite{clic_dataset} datasets. The ADAM optimizer is used with a fixed learning rate of $10^{-4}$. In the second stage (350 epochs for YOLOv3, 400 epochs for Faster R-CNN), equally sized random patches are taken from the larger VIMEO-90K~\cite{xue2019video_vimeo} dataset and a polynomial learning rate decay is introduced to allow fine-tuned learning.  

Training of the preview synthesis is done using a similar two-stage approach, but requires far fewer epochs to converge. Both the first and second stage are reduced to 50 epochs, with the learning rate decay accelerated to match the quicker training time. Notably, once the preview synthesis transform is trained, we can encode the remaining residual at any desired quality (and corresponding rate) by simply adjusting the QP value in the VTM configuration. 

\subsection{Base Layer Results}\label{subsec:base_results}

To evaluate the base-layer in a comparable way to previous benchmarks, we follow the approach set in~\cite{hyomin}. For the YOLOv3 model, we test our models on a subset of 5000 images from the COCO2014~\cite{COCO} dataset, and use the mean average precision (mAP) as measured at $50\%$ intersection over union (IoU), which is denoted as mAP@50, as our accuracy metric. Our results for YOLOv3 are compared with the best published settings of 3 previous scalable codecs~\cite{hyomin, pcs, ozyilkan2023learned}, refered to as Choi2022, Harell2022, and Ozyilkan2023. For completeness, we also include two traditional codecs, VVC-intra~\cite{vvc_std} and HEVC-intra~\cite{hevc_std_2019} (also known as BPG), and the learnable codec of~\cite{cheng2020image} to which we refer to as Cheng2020\footnote{For human-perception, results are borrowed from~\cite{hyomin} were compressed images were obtained using HM16-20~\cite{HM16.20}, VTM12.3~\cite{VTM12.3}, and CompressAI~\cite{begaint2020compressai} implementations for HEVC, VVC, and Cheng2020, respectively.}. For 
Faster R-CNN, 
we use the entire COCO2017~\cite{COCO} validation set (which also contains 5000 images), and report the average mAP over a range of IoU thresholds between $50-95\%$ with steps of $5\%$, which we simply denote mAP\footnote{The choice of two different mAP metrics is in order to better compare with~\cite{hyomin}, on which our base layer is based}. Available benchmarks for comparison here are Choi2022, Cheng2020 and the two traditional codecs VVC and HEVC.


\begin{figure}[htbp]
  \centering
  \includegraphics[width=0.9\linewidth]{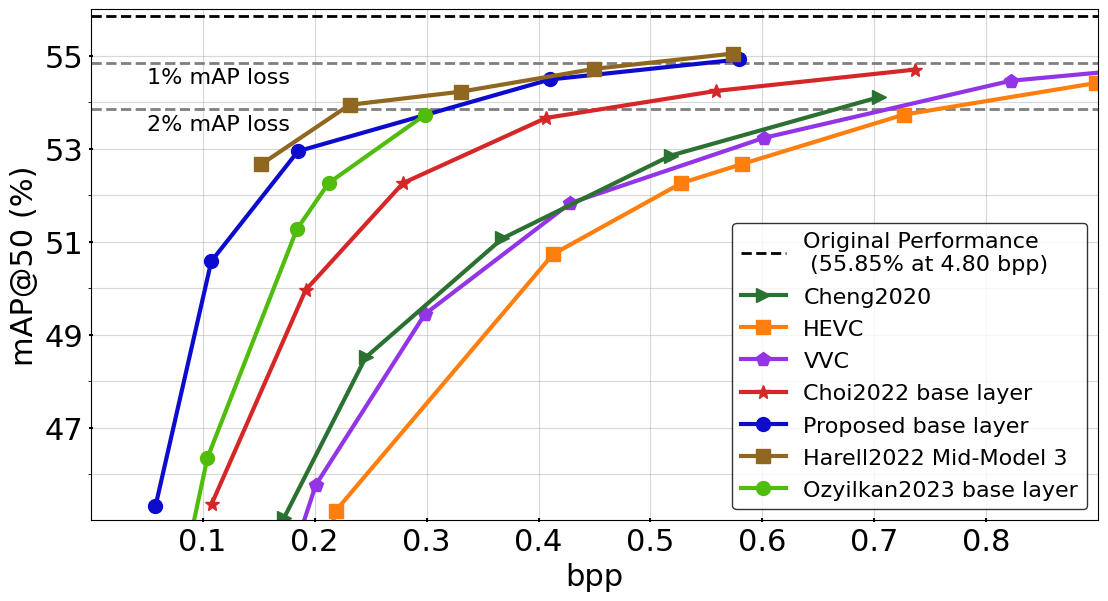}
  \caption{Base-layer object detection evaluation using YOLOv3.}
  \label{fig:base_yolo}
\end{figure}
\begin{figure}[htbp]
  \centering
  \includegraphics[width=0.9\linewidth]{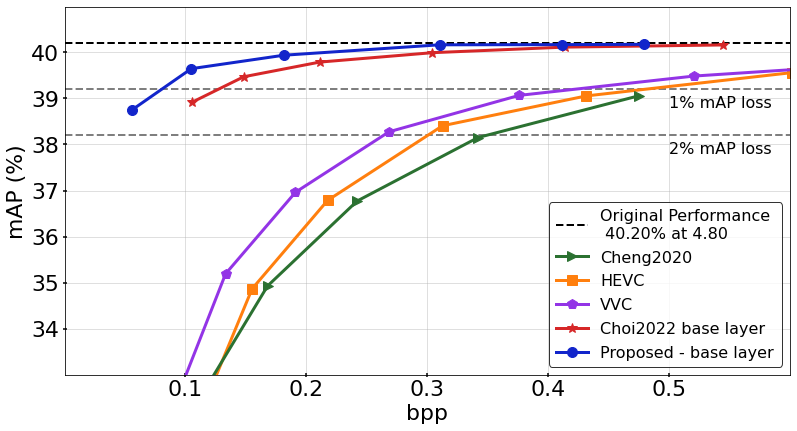}
  \caption{Base-layer object detection evaluation using Faster R-CNN.}
  \label{fig:base_faster}
\end{figure}

As 
seen in Fig.~\ref{fig:base_yolo}, our approach, which allows us to train the base-layer strictly based on task RD performance, achieves significant improvement over Choi2022 on which our base-layer is directly based. For example, for object detection using YOLOv3, our model suffers less than $1\%$ degradation in mAP@50\footnote{Compared to original YOLOv3 performance without compression.} at a rate of 0.58 bits per pixel (bpp), where~\cite{hyomin} was not able to achieve such accuracy even at 
rates of over 0.7 bpp. When compared with Harell2022, our model achieves slightly inferior performance, likely due to the use of deeper features in the training of Harell2022, which was proven to be superior, and can be incorporated in our model in future work. Compared with Ozyilkan2023, 
our model achieves superior performance, especially at lower rates, despite utilizing a simpler analysis transform. Our model's success 
relative to Ozyilkan2023 
is likely 
due to a combination of the simple entropy model used in~\cite{ozyilkan2023learned} and 
the latter being trained jointly. Comparing the results for Faster R-CNN, 
shown in Fig.~\ref{fig:base_faster}, we see that our model improves upon the previous SOTA of Choi2022, maintaining less than $1\%$ of reduction in mAP for rates as low as 0.1 bpp.

In order to numerically evaluate task RD performance we use a variant of the Bj\o{}ntegaard Delta metrics~\cite{Bjontegaard, vvc_ctc}, a well-established method for comparing rate-accuracy or rate-distortion curves. Because all base-layers here are used for CM, we replace the PSNR measurement, traditionally used in calculating BD-metrics, with mAP or mAP@50. The resulting metrics are calculated with respect to an anchor codec, for which we use Choi2022. BD-Rate estimates the average rate needed to achieve equivalent mAP as a percentage of the anchor's rate; BD-mAP estimates the improvement in mAP compared to the anchor when using equal bitrate.


\begin{table}[h!]
\centering
\setlength{\tabcolsep}{4pt}
\parbox{1\linewidth}{\centering\caption{Base-layer RD Performance Relative to Choi2022}\label{tbl:BD-base}}
\resizebox{1\linewidth}{!}{%
\begin{tabular}{@{}ccccc@{}}
\toprule
Model  &  \multicolumn{2}{c}{YOLOv3} & \multicolumn{2}{c}{ Faster R-CNN} \\  
& BD-Rate$[\%]$    & BD-mAP$[\%]$   &  BD-Rate$[\%]$   & BD-mAP$[\%]$  \\
\midrule
Proposed & --44.87 & \textbf{2.35} & \textbf{--36.6} & \textbf{0.27} \\
Choi2022~\cite{hyomin}  & 0 & 0 & 0 & 0\\
Harell2022~\cite{pcs}  & \textbf{--51.39} & 2.02 & - &  - \\
Ozyilkan2023~\cite{ozyilkan2023learned} & --24.04 & 1.52 & - &  - \\
\midrule
VVC~\cite{vvc_std}     & 58.93 & --3.15   & 239.4 & --2.33\\
HEVC~\cite{hevc_std_2019}      & 80.24 & --4.95   & 268.0 & --3.07 \\
Cheng2020~\cite{cheng2020image} & 51.88 & --2.91   & 305.8 & --3.62 \\
\end{tabular}
}
\end{table}

From Table~\ref{tbl:BD-base}, we see that our proposed method achieves significant rate savings compared to the base-layer of Choi2022 on which it is based, with a BD-Rate improvement of $44.9\%$ for YOLOv3 and $36.6\%$ for Faster R-CNN. In the case of YOLOv3, our work outperforms Ozyilkan2023 with BD-Rate savings of $24\%$, and slightly under-performs Harell2022 in terms of BD-Rate,  
but achieves higher BD-mAP than Harell2022. 
Also, the proposed method sets new SOTA results for Faster R-CNN.

\subsection{Enhancement Layer Results}\label{subsec:enh_results}

For brevity we present only RD curve for each of our task-models, and not the full combination of possible base and enhancement qualities, which can be seen in Appendix~\ref{app:ablation} of the supplemental material. In order to obtain a single curve for each task we limit the presented QP and $\lambda$ values while attempting to approximately match the mAP and PSNR values of Choi2022, and measure the required base and enhancement rates. By building our scalable codec this way, we can be certain that the comparison of the enhancement-layer is fair in terms of the base-layer performance. Points of similar PSNR correspond to similar mAP in the base-layer, and thus represent an equivalent use-case of the scalable codecs. 

\begin{figure}[htbp]
  \centering
  \includegraphics[width=0.9\linewidth]{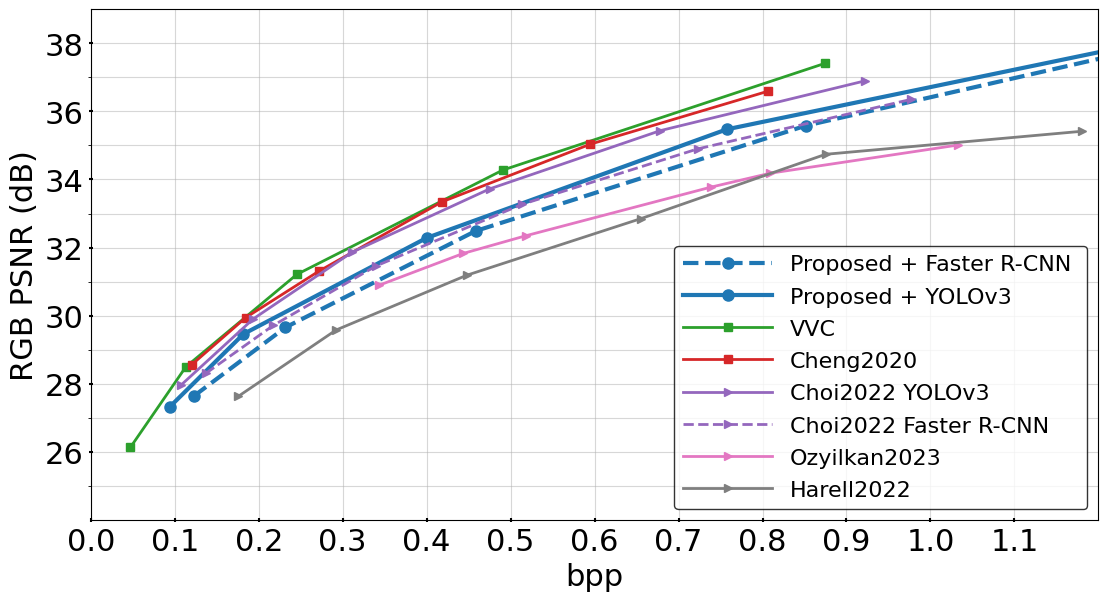}
  \caption{Benchmark comparison of the enhancement-layer of the proposed scalable codec. To maintain a fair comparison, we first choose the base-layer model to achieve similar task accuracy as~\cite{hyomin}, and then select an appropriate QP value for the enhancement to obtain comparable input PSNR.}
  \label{fig:enh_benchmark}
\end{figure}

The resulting models are evaluated on the Kodak~\cite{kodak_dataset} dataset alongside the same benchmarks used in the base-layer experiments\footnote{This time, as in~\cite{hyomin}, the benchamark results are borrowed directly from CompressAI~\cite{begaint2020compressai} where they are obtained using HM16-20~\cite{HM16.20}, VTM9.1~\cite{VTM9.1}, and CompressAI implementations for HEVC, VVC, and Cheng2020, respectively.}, as seen in Fig.~\ref{fig:enh_benchmark}. Immediately we notice that the proposed method achieves significantly better enhancement rate-distortion than both Harell2022 and Ozyilkan2023, which were comparable to it in the base-layer. Once again we quantify the differences between approaches using the BD-metrics, this time using the original PSNR to measure accuracy, and select VVC as the anchor, as seen in Table~\ref{tbl:BD-enh}. The proposed method with the YOLOv3 base-layer requires $23.2\%$ more bits than VVC, whereas Harell2022 and Ozyilkan2023 require $83.5\%$ and $62.6\%$ more bits, respectively. 

\begin{table}[h]
\centering
\setlength{\tabcolsep}{4pt}
\parbox{1\linewidth}{\centering\caption{Enh. Layer RD Performance Relative to VVC}\label{tbl:BD-enh}}
\resizebox{0.8\linewidth}{!}{%
\begin{tabular}{@{}ccc@{}}
\toprule
Model  &  \multicolumn{2}{c}{ Input Reconstruction} \\  
&  BD-Rate$[\%]$   & BD-PSNR[dB]  \\
\midrule
Proposed - Faster R-CNN                   & 38.52       & -1.40 \\
Proposed - YOLOv3                         & 23.19       & -0.84 \\
Choi2022~\cite{hyomin} - Faster R-CNN     & 30.60       & -1.14\\
Choi2022~\cite{hyomin} - YOLOv3           & 10.34       & -0.42\\
Harell2022~\cite{pcs}                     & 83.50       & -2.56\\
Ozyilkan2023~\cite{ozyilkan2023learned}   & 62.60       & -2.27\\
\midrule
VVC~\cite{vvc_std}                        & \textbf{0} & \textbf{0} \\
HEVC~\cite{hevc_std_2019}                 & 30.63       & -1.16 \\
Cheng2020~\cite{cheng2020image}               & 5.28       & -0.22 \\
\end{tabular}
}
\end{table}

When comparing with VVC, Cheng2020, and most importantly Choi2022, we see that the proposed method is slightly less efficient, requiring and average of $23.2\%$ and $38.5\%$ more rate when compared with VVC for the YOLOv3 and Faster R-CNN base-layers, respectively. For comparison, Cheng2020 requires a mere $5.3\%$ extra, and the Choi2022 scalable codec requires $10.34\%$ and $30.6\%$ corresponding to the same base-layers. Notably, the Choi2022 model corresponding to Faster R-CNN is actually a 3-layer model, and uses some rate to enable a second CV task, explaining why the YOLOv3 equivalent performs slightly better. 

Clearly, improving base-layer RD performance comes at a cost in the enhancement-layer, meaning the best solution depends on the relevant use case. 
One way to compare a scalable codec to a traditional codec, introduced in~\cite{Choi2022MMSP}, is to estimate the relative rate, $\mathcal{R}$, of a scalable codec compared to the traditional codec based on the frequency of use of the enhancement-layer: 
\begin{equation}
    \mathcal{R} = (1-f_H) \cdot\frac{R_{base}}{R} + f_H\cdot \frac{R_{base}+R_{enh}}{R}.
    \label{eq:fraction}
\end{equation}
Here $R_{base}$ and $R_{enh}$ correspond to the rates of the two layers of the scalable codec, while $R$ represents the rate of the traditional codec, and $f_H$ is the fraction of time where human viewing is required. When the relative rate is smaller than 1 the scalable codec is preferable. 
Unfortunately, Eq.~\ref{eq:fraction} is inadequate when comparing two scalable codecs because the denominator $R$ is not fixed for either codec. Instead we must use the following:
\begin{equation}
    \mathcal{R} = \frac{(1-f_H) \cdot R^{(c)}_{base} + f_H\cdot (R^{(c)}_{base}+R^{(c)}_{enh})}{(1-f_H) \cdot R^{(a)}_{base} + f_H\cdot (R^{(a)}_{base}+R^{(a)}_{enh})}.
    \label{eq:fraction_scalable}
\end{equation}
Where the superscript differentiates between the anchor (a) and the candidate codec (c). In order to simplify Eq.~\ref{eq:fraction_scalable} we assume a fixed ratio\footnote{In jointly trained codecs, this ratio is governed by the training loss and is often approximately fixed. In subsequent calculations, we estimated it empirically, by using the average measured rate.}, denoted $\rho$, between the rates of the enhancement and base layers of the codec in the denominator (the anchor). Doing this allows us to calculate enhancement frequency $f_H$ where the anchor and candidate codec offer equivalent performance using BD-Rates, denoted $BDR$ for both the base and enhancement layers.
\begin{equation}
    f_H^* = \frac{BDR_{base}}{BDR_{base}-\rho\cdot BDR_{total}},
\end{equation}
where both BD-Rates are calculated with respect to the anchor codec. If we choose an anchor which performs better on the enhancement-layer, then the candidate codec will be preferable in any scenario where human viewing is required less often then $f_H^*$, which we name the \emph{break even frequency}. We select Choi2022 as the anchor codec, resulting in a break-even frequency for our YOLOv3 model of $77.7\%$ which is significantly higher (better) than Harell2022 at $40.8\%$ and Ozyiklan2023 at $30.0\%$. For Faster R-CNN our model is preferable to Choi 2022 at any scenario where human viewing is needed less than $76.1\%$ of the time.

\section{Summary and Discussion}\label{sec:summary}

In this work we have presented VVC+M, a plug and play framework for converting any compression for machines approach into a scalable codec. Beginning with any ICM codec as our base-layer, we construct an enhancement-layer to perform input reconstruction. In our approach, we train a synthesis model to generate a preview image based on the machine codec's latent representation. Then we utilize the powerful compression capabilities of the inter-coding mode of VVC to encode the original image. We propose to do this by inserting the preview to to VVC's picture buffer, and detail a practical method for evaluating this approach using the publicly available VVC reference software VTM. Using our approach, the machine codec can be fully optimized for the automated analysis task, allowing for optimal performance for that scenario. Additionally, our approach allows for a complete decoupling between the qualities of the base and enhancement layers by simply changing the QP value used in VVC. Another important aspect of our framework is that it can be modified to be used with any ICM model as well as any inter-coding approach, including highly efficient codecs designed specifically for computationally limited edge devices.

We demonstrate our approach by improving the base-layer of~\cite{hyomin}, which until recently represented the state-of-the-art ICM model for object detection. Our codec shows significant rate savings in the base-layer of up to $44.9\%$, approaching or improving SOTA compression for this task. As can be expected, the improvement of base-layer performance comes at the cost of some degradation in the enhancement-layer rate-distortion. In order to evaluate the overall benefit of our approach, we propose a method of calculating the \emph{relative rate} of two scalable codecs, based on the frequency of use of the enhancement-layer. We find that our model outperforms all comparable codecs whenever the full input reconstruction is needed less than $76$-$77\%$ of the time.  

\begin{small}  
\bibliographystyle{IEEEtran}
\bibliography{icme2023template}
\end{small}
\newpage
\onecolumn

\appendix

\begin{center}
\bf
SUPPLEMENT TO

VVC+M: PLUG AND PLAY SCALABLE IMAGE CODING FOR HUMANS AND MACHINE
\end{center}

\section{Base Layer Details}\label{app:base_layer}
\begin{figure}[htbp]
  \centering
  \includegraphics[width=0.9\linewidth]{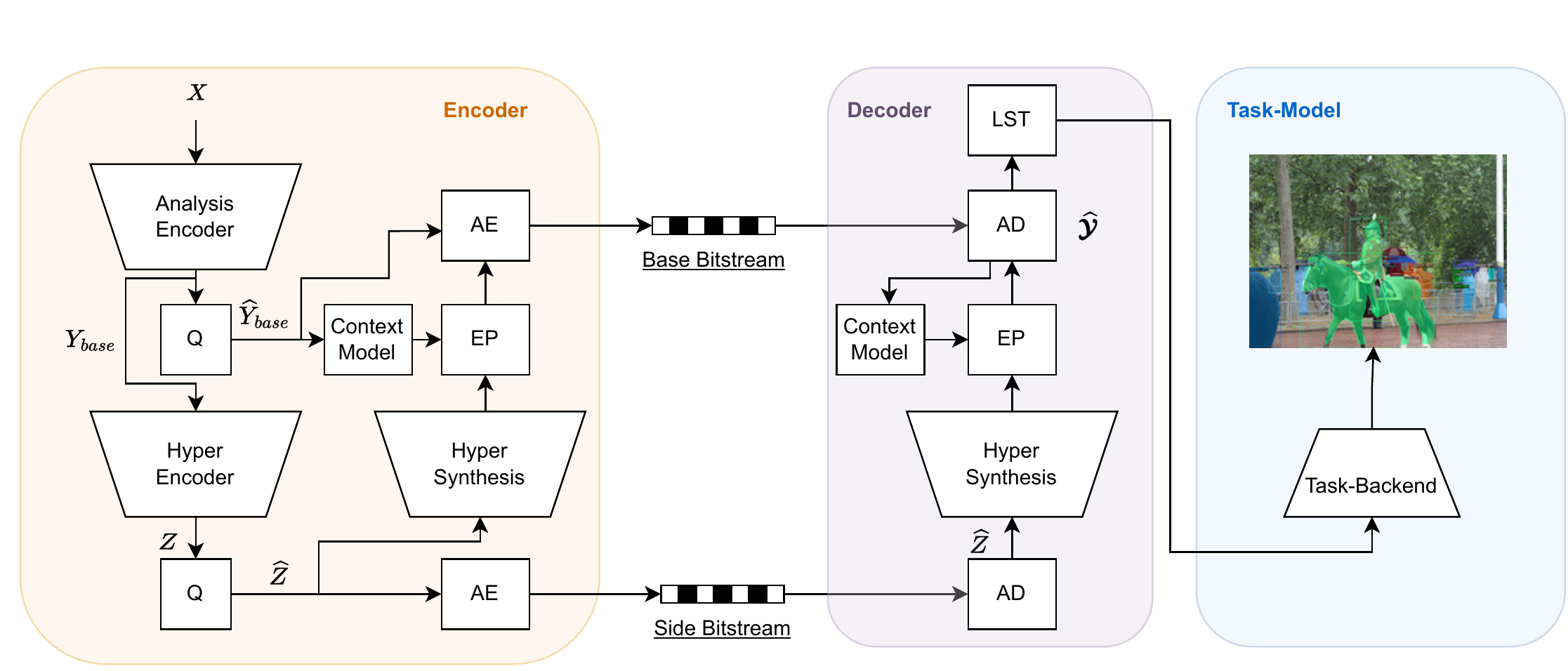}
  \caption{Base-layer encoder and decoder implementation -  Q stands for quantization; EP is the entropy parameter estimation blocks; AE and AD are arithmetic encoding and decoding, respectively; and LST is the latent-space transform. }
  \label{fig:base-arch}
\end{figure}

As mentioned in Section~\ref{subsec:proposed_base}, our base-layer implementation is made as similar as possible to~\cite{hyomin}, and can be seen in Figure~\ref{fig:base-arch}.
On the encoder side, an input $X$ is passed to an analysis transform $g_a^{base}$ which is comprised of residual convolutional blocks and downsampling blocks, followed by leaky ReLU and generalized divisive normalization (GDN) activations~\cite{gdn}. The resulting latent representation $Y_{base}$ is then further analyzed by a fully convolutional network $h_a$ to create the side information $Z$, which is quantized to $\widehat{Z}$. Next the side information itself is encoded using an entropy bottleneck model~\cite{balle2018variational}, followed by quantization and an arithmetic encoder. The quantized side information is then passed to a hyper-synthesis network, similarly to~\cite{minnen2018joint,cheng2020image}, to produce initial parameter estimates for the distribution of the latent representation $Y_{base}$ (a Gaussian distribution in~\cite{hyomin, minnen2018joint} and our work, and a Gaussian mixture model in~\cite{cheng2020image}). Alongside the initial estimates, an autoregressive context model~\cite{minnen2018joint} is employed to obtain further parameter estimates directly from previously encoded elements of $Y_{base}$. An entropy parameter estimation block, made of several $1\times1$ convolutional blocks is then used to merge the two density parameter estimates and calculate the necessary probability estimates for an arithmetic encoding of the quantized representation $\widehat{Y}_{base}$. 

On the decoder side, the side information bitstream is analysed first, recreating $\widehat{Z}$ using an arithmetic decoder. Using exact copies of the hyper-synthesis network and the autoregressive context model, the side information is used to sequentially estimate and the density of each element $\widehat{Y}_{base}$ subsequently reconstruct it from the main bitstream using a second arithmetic decoder. Having recreated the latent representation, we then transform it using a latent-space transform (LST), which is identical to that presented in~\cite{hyomin}, to match the dimensions of some feature layer $F$ in the task model. The LST itself is comprised of similar layers to those of the synthesis transform in~\cite{minnen2018joint, cheng2020image}, including upsampling and residual blocks using inverse GDN activations.  Finally, the reconstructed features, $\widehat{F}$ are passed to the remaining layers of the task model to obtain the desired task labels.

\section{Enhancement Layer Ablation Study}\label{app:ablation}
\begin{figure}[htbp]
    \centering
        \includegraphics[width=0.8\textwidth]{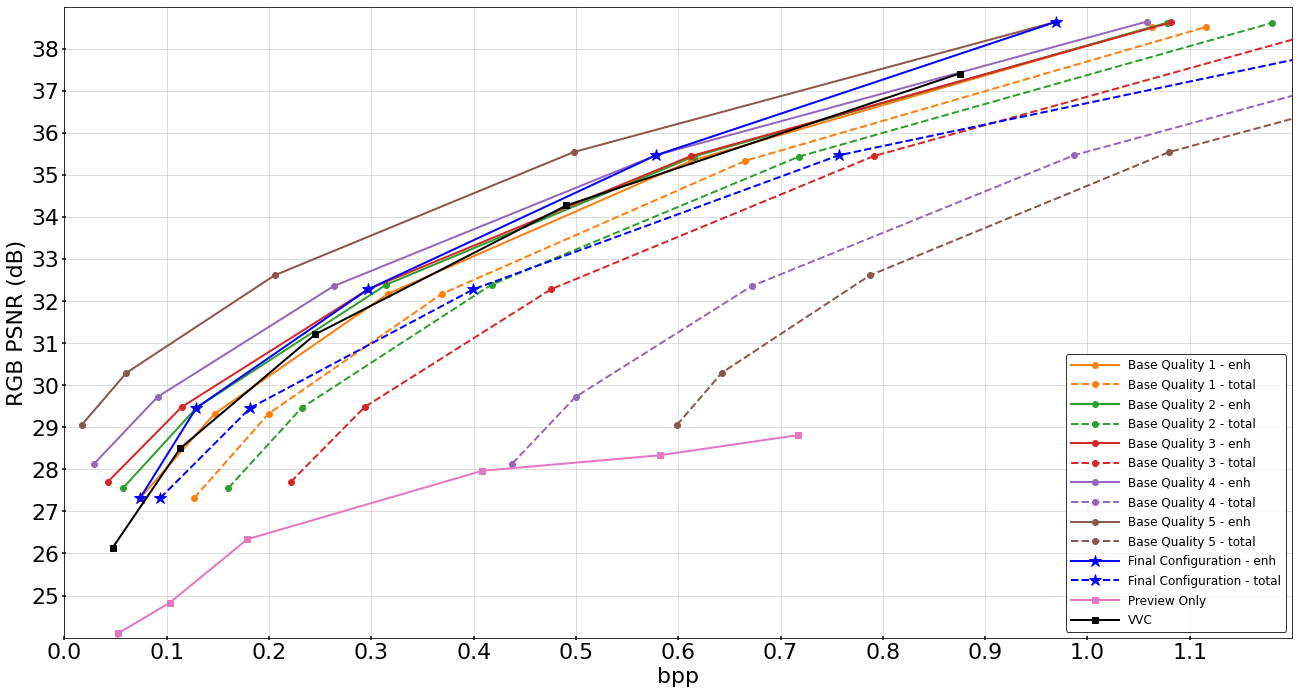}
        \caption{Rate-distortion curves for all combinations of YOLOv3 base-layer models and enhancement-layer QP values. }
        \label{fig:ablation_yolo}
\end{figure}

\begin{figure}[htbp]
        \centering
        \includegraphics[width=0.8\textwidth]{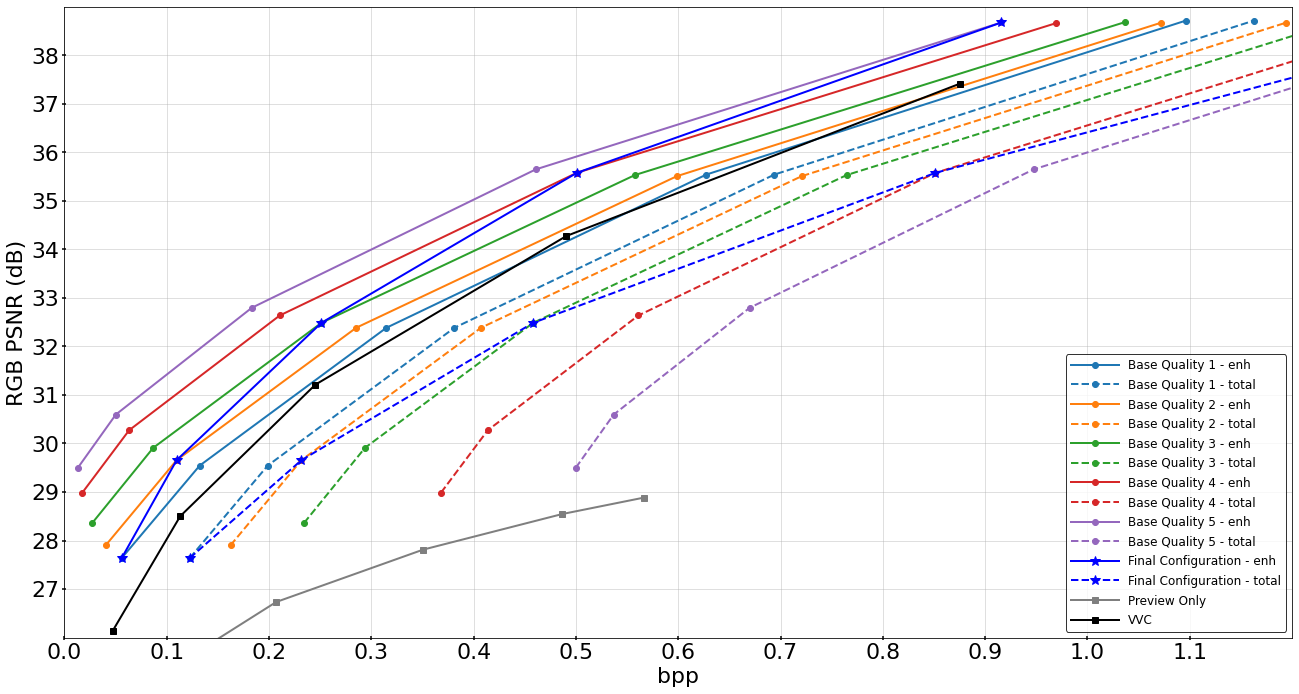}
        \caption{Rate-distortion curves for all combinations of Faster R-CNN base-layer models and enhancement-layer QP values.}
        \label{fig:ablation_faster}
 \end{figure}

As explained in Section~\ref{subsec:proposed_enh}, our proposed approach allows for controlling the base-layer quality separately from that of the enhancement-layer. Thus, we are able to produce an entire RD curve for each base-layer model, simply by changing the QP parameter used in VVC, as can be seen in Figures~\ref{fig:ablation_yolo}, \ref{fig:ablation_faster}. Furthermore, it is important to evaluate whether or not the enhancement-layer is preferable to simply re-transmitting the entire image. To show this comparison, we also include RD curves where only the rate of the enhancement-layer is measured, alongside the VVC-intra baseline. Additionally, we also include the figure of our final configuration used for comparison with other benchmarks in Section~\ref{subsec:enh_results}, and the RD-curve for the preview only.

Observing the results in both figures, we clearly see that in every configuration, our enhancement-layer is slightly better than the baseline of simply using VVC-intra. Interestingly, we see that as currently constructed base representation is not very well-suited for input reconstruction, leading to poor RD-curve for the preview. Furthermore, we can see that the rate used on the base representation does not get fully utilized by the enhancement-layer (as is generally the case). One possible solution for this problem would be to include a small input reconstruction loss while training the base-layer, weighted far more lightly than the task loss. An alternative solution may be to train the preview layers using a proxy loss for the VVC-inter coding, in place of the current MSE, similar to the approach taken in~\cite{guleryuz2022sandwiched}.

  \newpage
\section{Example VTM Configuration File}\label{app:cfg}

Below is an example configuration file corresponding to \texttt{QP = 28}. Please note that some lines have been split to multiple lines to prevent overflow onto the margins. Such lines are highlighted in the color blue, and must be used as a single line when utilized in practice to configure VTM.

\begingroup
\setlength\parindent{0pt}
\footnotesize
 \ttfamily
 \noindent
    \#======== File I/O =====================
    
BitstreamFile                 : str.bin 

ReconFile                     : rec.yuv

\medskip
\#======== Profile ================

Profile                       : auto

\medskip

\#======== Unit definition ================

MaxCUWidth                    : 64          \# Maximum coding unit width in pixel

MaxCUHeight                   : 64          \# Maximum coding unit height in pixel

\medskip
\#======== Coding Structure =============

IntraPeriod                   : -1          \# Period of I-Frame ( -1 = only first)

DecodingRefreshType           : 0           \# Random Accesss 0:none, 1:CRA, 2:IDR, 3:Recovery Point SEI

GOPSize                       : 1           \# GOP Size (number of B slice = GOPSize-1)

\medskip
IntraQPOffset                 : -28

\textcolor{blue}{LambdaFromQpEnable            : 1           \# see JCTVC-X0038 for suitable parameters for IntraQPOffset, QPoffset, QPOffsetModelOff, QPOffsetModelScale when enabled}

\textcolor{blue}{\#        Type POC QPoffset QPOffsetModelOff QPOffsetModelScale CbQPoffset CrQPoffset QPfactor tcOffsetDiv2 betaOffsetDiv2 CbTcOffsetDiv2 CbBetaOffsetDiv2 CrTcOffsetDiv2 CrBetaOffsetDiv2 temporal\_id \#ref\_pics\_active\_L0 \#ref\_pics\_L0   reference\_pictures\_L0 \#ref\_pics\_active\_L1 \#ref\_pics\_L1   reference\_pictures\_L1}

Frame1:  P    1   0       -6.5               0.2590         0          0          1.0      0            0               0             0                 0               0               0             1                1         1                      0                   0

\medskip

\#=========== Motion Search =============

FastSearch                    : 1           \# 0:Full search  1:TZ search

SearchRange                   : 64          \# (0: Search range is a Full frame)

BipredSearchRange             : 4           \# Search range for bi-prediction refinement

HadamardME                    : 1           \# Use of hadamard measure for fractional ME

FEN                           : 1           \# Fast encoder decision

FDM                           : 1           \# Fast Decision for Merge RD cost

\medskip

\#======== Quantization =============

QP                            : 28           \# CU-based multi-QP optimization

MaxCuDQPSubdiv                : 0           \# Maximum subdiv for CU luma Qp adjustment

DeltaQpRD                     : 0           \# Slice-based multi-QP optimization

RDOQ                          : 1           \# RDOQ

RDOQTS                        : 1           \# RDOQ for transform skip

\medskip

\#=========== Deblock Filter ============

\textcolor{blue}{LoopFilterOffsetInPPS         : 1           \# Dbl params: 0=varying params in SliceHeader, param = base\_param + GOP\_offset\_param; 1 (default) =constant params in PPS, param = base\_param)}

LoopFilterDisable             : 0           \# Disable deblocking filter (0=Filter, 1=No Filter)

LoopFilterBetaOffset\_div2     : 0           \# base\_param: -12 ~ 12

LoopFilterTcOffset\_div2       : 0           \# base\_param: -12 ~ 12

LoopFilterCbBetaOffset\_div2   : 0           \# base\_param: -12 ~ 12

LoopFilterCbTcOffset\_div2     : 0           \# base\_param: -12 ~ 12

LoopFilterCrBetaOffset\_div2   : 0           \# base\_param: -12 ~ 12

LoopFilterCrTcOffset\_div2     : 0           \# base\_param: -12 ~ 12

\textcolor{blue}{DeblockingFilterMetric        : 0           \# blockiness metric (automatically configures deblocking parameters in bitstream). Applies slice-level loop filter offsets (LoopFilterOffsetInPPS and LoopFilterDisable must be 0)}

\medskip

\#=========== Misc. ============

InternalBitDepth              : 10          \# codec operating bit-depth

\medskip

\#=========== Coding Tools =================

SAO                           : 1           \# Sample adaptive offset  (0: OFF, 1: ON)

TransformSkip                 : 1           \# Transform skipping (0: OFF, 1: ON)

TransformSkipFast             : 1           \# Fast Transform skipping (0: OFF, 1: ON)

TransformSkipLog2MaxSize      : 5

SAOLcuBoundary                : 0           \# SAOLcuBoundary using non-deblocked pixels (0: OFF, 1: ON)

\medskip

\#=========== TemporalFilter =================

TemporalFilter                : 0           \# Enable/disable GOP Based Temporal Filter

TemporalFilterFutureReference : 0           \# Enable/disable reading future frames

TemporalFilterStrengthFrame4  : 0.4         \# Enable filter at every 4th frame with strength

\medskip

\#============ Rate Control ======================

RateControl                         : 0                \# Rate control: enable rate control

TargetBitrate                       : 1000000          \# Rate control: target bitrate, in bps

\textcolor{blue}{KeepHierarchicalBit                 : 2                \# Rate control: 0: equal bit allocation; 1: fixed ratio bit allocation; 2: adaptive ratio bit allocation}

LCULevelRateControl                 : 1                \# Rate control: 1: LCU level RC; 0: picture level RC

RCLCUSeparateModel                  : 1                \# Rate control: use LCU level separate R-lambda model

InitialQP                           : 0                \# Rate control: initial QP

RCForceIntraQP                      : 0                \# Rate control: force intra QP to be equal to initial QP

\medskip

\#============ VTM settings ======================

SEIDecodedPictureHash               : 0

CbQpOffset                          : 0

CrQpOffset                          : 0

SameCQPTablesForAllChroma           : 1

QpInValCb                           : 17 22 34 42

QpOutValCb                          : 17 23 35 39

ReWriteParamSets                    : 1

\medskip

\#============ NEXT ====================
\medskip

\# General

CTUSize                      : 128

LCTUFast                     : 1

DualITree                    : 1      \# separate partitioning of luma and chroma channels for I-slices

MinQTLumaISlice              : 8

MinQTChromaISliceInChromaSamples: 4      \# minimum QT size in chroma samples for chroma separate tree

MinQTNonISlice               : 8

MaxMTTHierarchyDepth         : 3

MaxMTTHierarchyDepthISliceL  : 3

MaxMTTHierarchyDepthISliceC  : 3

MTS                          : 1

MTSIntraMaxCand              : 3

MTSInterMaxCand              : 4

SBT                          : 1

ISP                          : 1

Affine                       : 1

SbTMVP                       : 1

MaxNumMergeCand              : 6

LMChroma                     : 1      \# use CCLM only

DepQuant                     : 1

IMV                          : 1

ALF                          : 1

CIIP                         : 1

IBC                          : 0      \# turned off in CTC

AllowDisFracMMVD             : 1

AffineAmvr                   : 0

LMCSEnable                   : 1      \# LMCS: 0: disable, 1:enable

LMCSSignalType               : 0      \# Input signal type: 0:SDR, 1:HDR-PQ, 2:HDR-HLG

LMCSUpdateCtrl               : 2      \# LMCS model update control: 0:RA, 1:AI, 2:LDB/LDP

LMCSOffset                   : 1      \# chroma residual scaling offset

MRL                          : 1

MIP                          : 0

JointCbCr                    : 1      \# joint coding of chroma residuals (if available): 0: disable, 1: enable

PROF                         : 1

ChromaTS                     : 1

\medskip

\# Fast tools

PBIntraFast                  : 1

ISPFast                      : 0

FastMrg                      : 1

AMaxBT                       : 1

FastMIP                      : 0

FastLocalDualTreeMode        : 2

\medskip

\# Encoder optimization tools

AffineAmvrEncOpt             : 0

MmvdDisNum                   : 6

ALFAllowPredefinedFilters    : 1

ALFStrengthTargetLuma        : 1.0

ALFStrengthTargetChroma      : 1.0

CCALFStrengthTarget          : 1.0

\#\#\# DO NOT ADD ANYTHING BELOW THIS LINE \#\#\#
\#\#\# DO NOT DELETE THE EMPTY LINE BELOW \#\#\#

\endgroup





\end{document}